\documentclass[conference]{IEEEtran}
\IEEEoverridecommandlockouts
\usepackage{cite}
\usepackage{amsmath,amssymb,amsfonts}
\usepackage{algorithmic}
\usepackage{graphicx}
\usepackage{textcomp}
\usepackage{xcolor}
\usepackage{etex}
\usepackage{booktabs}
\usepackage[utf8]{inputenc}
\usepackage[ruled]{algorithm2e}
\usepackage[T1]{fontenc}
\usepackage{microtype}
\usepackage{paralist}
\usepackage{tabularx}
\usepackage{balance}
\usepackage{multicol}
\usepackage{multirow}
\usepackage{pbox}
\usepackage{colortbl}
\usepackage{pifont}
\usepackage{xspace}
\usepackage{xurl}
\usepackage{tikz}
\usepackage{rotating}
\usepackage{float}
\usepackage{subcaption}
\usepackage{listings}
\usepackage{makecell} 
\usepackage{hhline}
\usepackage{hyperref}
\usepackage{verbatim}
\usepackage{makecell}
\usepackage[flushleft]{threeparttable}
\usepackage{balance}
\usepackage{float}
\usepackage{inconsolata}
\floatstyle{plaintop}
\restylefloat{table}

\newcolumntype{C}[1]{>{\centering\arraybackslash}m{#1}}
\newcolumntype{G}{>{\centering\arraybackslash}m{0.7cm}}
\newcolumntype{H}{>{\centering\arraybackslash}p{1.5cm}}
\newcolumntype{I}{>{\centering\arraybackslash}p{2cm}}
\newcolumntype{Y}{>{\centering\arraybackslash}p{2.7cm}}

\usepackage{multirow}
\usepackage{hhline}
\usepackage{listings}
\usepackage{array}
\usepackage{pifont}
\usepackage{enumerate}
\usepackage[shortlabels]{enumitem}
\usepackage{soul}

\lstdefinestyle{jsonstyle}{
    basicstyle=\ttfamily\small,
    keywordstyle=\color{blue},
    stringstyle=\color{red},
    showstringspaces=false,
    breaklines=true,
    moredelim=[is][\color{red}]{<HL>}{</HL>},
    moredelim=[is][\color{blue}]{<HL2>}{</HL2>}
}
\lstdefinestyle{yaml}{
     basicstyle=\color{blue}\footnotesize,
     rulecolor=\color{black},
     string=[s]{'}{'},
     stringstyle=\color{blue},
     comment=[l]{:},
     commentstyle=\color{black},
 }

\newcommand{\ie}{\emph{i.e.,}\xspace}
\newcommand{\eg}{\emph{e.g.,}\xspace}

\newcommand{\etal}{\emph{et~al.}\xspace}

\newcommand{\figref}[1]{Fig.~\ref{#1}\xspace}
\newcommand{\listref}[1]{Listing~\ref{#1}\xspace}
\newcommand{\tabref}[1]{Table~\ref{#1}\xspace}
\newcommand{\approach}{\textsc{FLOWSTEP}\xspace}
\newcommand{\rqa}{How effective is \approach in generating accurate recommendations for GitHub Actions workflows?}
\newcommand{\rqb}{What is the impact of different model configurations within the \approach framework on its final performance?}

\newboolean{showcomments}

\setboolean{showcomments}{true}

\ifthenelse{\boolean{showcomments}}
  {\newcommand{\nb}[2]{
    \fbox{\bfseries\sffamily\scriptsize#1}
    {\sf\small$\blacktriangleright$\textit{#2}$\blacktriangleleft$}
   }
  }
  {\newcommand{\nb}[2]{}
  }

\def\BibTeX{{\rm B\kern-.05em{\sc i\kern-.025em b}\kern-.08em
    T\kern-.1667em\lower.7ex\hbox{E}\kern-.125emX}}
\begin{document}

%\title{Has GitHub Actions Workflow Rerun Been Abused? An Empirical Study}

\title{FLOWSTEP: A Framework for Learning to Predict GitHub Actions Workflow STEPs}

%\author{~\\Anonymous Authors\\Anonymous Institution\\~\\}
\author{\IEEEauthorblockN{Jiangnan Huang}
\IEEEauthorblockA{\textit{Radboud University} \\
Nijmegen, The Netherlands\\
jiangnan.huang@ru.nl}
\and
\IEEEauthorblockN{Bin Lin}
\IEEEauthorblockA{\textit{Hangzhou Dianzi University} \\
Hangzhou, China\\
b.lin@hdu.edu.cn}
\and
\IEEEauthorblockN{Yutaro Kashiwa}
\IEEEauthorblockA{
\textit{Nara Institute of}\\
\textit{Science and Technology}\\
Nara, Japan\\
yutaro.kashiwa@is.naist.jp
}
}

\maketitle
\thispagestyle{plain}
\pagestyle{plain}

\begin{abstract}
%Context
Continuous Integration and Delivery (CI/CD) are fundamental to modern software development, yet configuring pipelines such as GitHub Actions (GHA) workflows remains complex and error-prone.
%Problem
Recent work has explored Transformer-based models for GHA workflow recommendation, but existing approaches typically model workflow steps as a single homogeneous generation task. This formulation overlooks a key distinction: reusable \textit{Actions} are structured, ecosystem-level components, whereas project-specific \textit{Scripts} encode local build, test, and deployment semantics. Consequently, current models struggle to capture the different forms of knowledge required for accurate workflow completion.
%Recent studies have applied Transformer-based models to GHA workflow recommendation; however, these approaches often suffer from limited accuracy and struggle to capture heterogeneous workflow elements, such as \textit{Action} and \textit{Script} generation. 
%Moreover, current evaluation metrics such as accuracy often overlook the semantics of generated scripts, offering only a shallow assessment of their practical utility.
%
In this work, we present \approach, a framework for GHA workflow step completion that explicitly distinguishes between Action and Script generation. \approach combines large language models (LLMs) with contrastive learning to better capture semantic relationships among workflow contexts, Actions, and Scripts. 
%In this work, we present \approach, a framework for GHA workflow step completion that distinguishes \textit{Action} and \textit{Script} generation and combines large language models (LLMs) with contrastive learning to better capture workflow semantics. 
%Results
Evaluated on established benchmarks, \approach improves end-to-end workflow step prediction accuracy from 38.23\% to 51.09\%. Beyond prediction accuracy, we further study whether LLM-as-a-judge can provide a more semantic evaluation of generated workflow steps. By comparing LLM-based assessments with human annotations, we find that the best-performing model configuration reaches substantial agreement with human ground truth (Krippendorff's $\alpha = 0.701$), suggesting that LLM-as-a-judge can serve as a scalable complement to manual evaluation.
%Evaluating on established benchmarks, \approach improves end-to-end workflow step prediction accuracy from 38.23\% to 51.09\%. We also investigate the feasibility of using LLM-as-a-judge to evaluate generated workflow steps by comparing its assessments with human annotations. The results show strong agreement between LLM-based evaluation and human judgment, suggesting that LLM-as-a-judge can serve as a scalable complement to manual evaluation.

\end{abstract}


%%%%%%%%%%%%%%%%%%%%%%%
\section{Introduction}
\label{sec:intro}

%GitHub has grown into the largest ecosystem for collaborative software development, connecting more than 180 million developers and hosting over 630 million repositories as of 2025~\cite{github2025}. %Over time, this platform has reshaped how software teams collaborate and share code~\cite{inproceedings,10.1145/2568225.2568260}
%As software projects grow in scale, Continuous Integration and Delivery (CI/CD) has shifted from an optional practice to a fundamental necessity for automating building, testing, and deployment~\cite{rostami2023usage}. 
% As projects on GitHub became increasingly complex, developers began adopting continuous integration and delivery (CI/CD) services to automate essential development tasks such as building, testing, and deployment~\cite{rostami2023usage}.
%To natively support these activities, GitHub launched GitHub Actions (GHA)~\cite{github-actions} in November 2019, offering an integrated CI/CD environment directly within the platform. Within only 18 months, GHA surpassed Travis CI to become the most widely used automation framework on GitHub~\cite{9825792}. Its convenience and ecosystem of reusable components (\textit{Actions}) have made workflow automation more accessible than ever.

GitHub has become the largest collaborative software ecosystem, connecting more than 180 million developers and hosting over 630 million repositories as of 2025~\cite{github2025}. As projects grow, Continuous Integration and Delivery (CI/CD) has become essential for automating building, testing, and deployment~\cite{rostami2023usage}. To natively support these activities, GitHub launched GitHub Actions (GHA)~\cite{github-actions} in November 2019, offering an integrated CI/CD environment directly within the platform. Within only 18 months, GHA surpassed Travis CI to become the most widely used automation framework on GitHub~\cite{9825792}. Its convenience and ecosystem of reusable components (\textit{Actions}) have made workflow automation more accessible than ever.

Despite these advantages, configuring CI/CD pipelines remains challenging. Prior studies have repeatedly shown that developers struggle with pipeline setup, maintenance, and debugging~\cite{10.1109/MS.2015.27,Empirical_Zampetti}. In a survey by Saroar \etal~\cite{saroar2023developersperceptiongithubactions}, about 60\% of 90 participants reported difficulties when automating workflows using GHA, underscoring an urgent need for intelligent authoring support.
%Despite these advantages, configuring and maintaining CI/CD pipelines remains a challenging endeavor. Prior studies have repeatedly shown that developers struggle with pipeline setup, maintenance, and debugging~\cite{10.1109/MS.2015.27,10.1145/3106237.3106270,saroar2023developersperceptiongithubactions,10.1145/3524842.3527948,Empirical_Zampetti}. Although CI/CD improves software quality, productivity, and release speed~\cite{CHEN201772,10.1145/2786805.2786850}, its successful adoption often depends on project-specific environments and technologies~\cite{10.1145/3106237.3106270,10.1145/3524842.3527948}. For instance, pipelines may need to accommodate multiple operating systems or embedded devices, requiring extensive customization. In a survey by Soroar \etal~\cite{saroar2023developersperceptiongithubactions}, about 60\% of 90 participants reported difficulties when automating workflows using GitHub Actions—illustrating the persistent complexity of CI/CD authoring.
While automated code assistance has seen breakthroughs in general programming tasks (\eg code completion
% This growing need for intelligent support has parallels in the broader field of automated code assistance. Machine learning has been widely applied to automated code assistance, including code completion
~\cite{Ciniselli_2021}, assertion generation~\cite{Watson_2020}, automated repair of vulnerabilities and bugs~\cite{Chen_2023,Michael_2022}), workflow authoring introduces unique challenges. A CI/CD pipeline interleaves reusable \textit{Actions} with free-form scripting and natural-language documentation, while embedding highly context-specific details such as directory paths or download URLs. These characteristics make workflow recommendation significantly harder than traditional code completion.

The rapid adoption of GHA has motivated learning-based workflow generation. Transformer models and LLMs have been used to synthesize or recommend workflow steps~\cite{10.1145/3597503.3623351,10.1145/3664476.3664497}, but existing approaches often treat steps as homogeneous generation targets. This overlooks the different prediction strategies required by \textit{Actions} and shell \textit{Scripts}: \textit{Actions} require selecting a reusable identifier from the GHA ecosystem, whereas \textit{Scripts} require generating project-specific shell commands from workflow context. Our analysis shows that this mismatch contributes to prediction errors and reduces robustness across heterogeneous workflow steps.

In this work, we address the problem of recommending the next step in GHA workflows, accounting for the heterogeneous nature of workflow steps. To this end, we propose \approach, a multi-step prediction framework comprising (i) a binary classifier that predicts whether the next step is an \textit{Action} or a \textit{Script}, (ii) a RoBERTa model with contrastive learning that predicts the specific \textit{Action}, and (iii) an LLM that generates the step when it is a \textit{Script}. For comparability with prior work, we use the GHA workflow corpus by Decan \etal~\cite{actiondata}, further processed by Mastropaolo \etal~\cite{10.1145/3597503.3623351} into two variants: a raw version preserving original content and an abstracted version replacing project-specific tokens (e.g., paths) with placeholders while retaining commands and tool options. We evaluate \approach on both variants to assess robustness across data granularities. The next step is predicted from the preceding workflow context concatenated with the step’s descriptive name, reflecting real workflow authoring scenarios.

The main contributions of our study are as follows:
\begin{itemize}

    \item We formulate GHA workflow step recommendation as a heterogeneous prediction problem and present \approach, a step-type-aware framework that  differentiates reusable \textit{Actions} from project-specific \textit{Scripts}, improving prediction accuracy from 38.23\% to 51.09\% (raw data).
    %\item We present a GHA workflow step recommendation framework, \approach, which explicitly differentiates between \textit{Actions} and \textit{Scripts}, improving step prediction accuracy from 38.23\% to 51.09\% (raw data).
    \item We conduct a comprehensive evaluation of different model configurations for \approach. Combining a RoBERTa-based model for \textit{Action} prediction with Claude-Sonnet-4.5 for \textit{Script} generation yields the best performance.
    \item We investigate the feasibility of using \textit{LLM-as-a-judge} for evaluating generated workflow \textit{Scripts} by comparing LLM judgments with human annotations on 350 sampled predictions, analyzing their alignment and reliability.
    \item We make publicly available our approach scripts, checkpoints,  predictions, and the used datasets.
    
\end{itemize}

%%%%%%%%%%%%%%%%%%%%%%%
\section{Background}
\label{sec:background}

%In this section, we introduce background knowledge of GHA, explain workflow rerun and job rerun process, and provide two real-world examples of rerun waste to motivate our study.

\subsection{GitHub Actions Workflow}

A GHA workflow is a YAML configuration file stored in the \texttt{.github/workflows} directory of a repository. It defines automated tasks that support CI/CD activities such as building, testing, and deployment. Workflows are triggered by predefined \textit{events} (e.g., \texttt{push}, \texttt{pull\_request}, or scheduled \texttt{cron} jobs)~\cite{ghawf}. An example can be found in \listref{lst:yaml}, which defines a GHA workflow named \texttt{Maven test}\footnote{The example is adapted from a randomly picked GHA workflow, available on \url{https://github.com/abel533/Mapper/blob/master/.github/workflows/test.yml}}. 

\begin{lstlisting}[style=yaml,caption={Example of a GHA workflow file},captionpos=b,label={lst:yaml},frame=tb]
name: Maven test
on:
  pull_request:
    types: [ opened, reopened, edited ]
  push:
    branches: [ develop, master ]
jobs:
  mvn_verify:
    runs-on: ubuntu-latest
    steps:
      - uses: actions/checkout@v3
      - name: Set up Maven Central Repository
        uses: actions/setup-java@v2
        with:
          java-version: '17'
          distribution: 'adopt'
      - name: Run the Maven verify phase
        run: mvn --batch-mode -P dev test
\end{lstlisting}

Each workflow contains one or more \textit{jobs}, representing independent execution units that run on GitHub-hosted or self-hosted \textit{runners}. Jobs may run concurrently or sequentially, and they remain isolated unless they explicitly share artifacts or caches. Each job consists of a series of \textit{steps}, the smallest executable elements in a workflow. Steps can be defined either with the \texttt{run} keyword to execute shell commands (\textit{Scripts}) directly within the runner (e.g., \texttt{gradle build}) or with the \texttt{uses} keyword to invoke reusable \textit{Actions}~\cite{ghaaction}. \textit{Script}-based steps are preferred for simple or project-specific tasks, whereas \textit{Actions} encapsulate recurring operations and can be versioned for reproducibility (e.g., \texttt{actions/checkout@v3}). Through this declarative structure, GHA provides a unified mechanism to automate CI/CD pipelines directly within GitHub repositories.

\subsection{Transformers and Contrastive Learning}
The evolution of Transformer architectures~\cite{10.5555/3295222.3295349} and encoders like BERT~\cite{devlin2019bert} has enabled models to learn deep contextual representations of text and code. RoBERTa~\cite{liu2019roberta}, a refinement of BERT, employs dynamic masking and achieves more stable and robust performance across various Natural Language Processing tasks.
% The success of deep learning in Natural Language Processing (NLP) largely stems from the introduction of Transformer architectures~\cite{10.5555/3295222.3295349}, which rely on self-attention mechanisms to capture dependencies among sequence elements without recurrence, enabling large-scale parallel training. Building on this foundation, BERT~\cite{devlin2019bert} introduced a bidirectional Transformer encoder that learns contextual representations through Masked Language Modeling (MLM) and Next Sentence Prediction (NSP). RoBERTa~\cite{liu2019roberta}, a refinement of BERT, removes the NSP objective and employs dynamic masking, leading to more stable and robust performance across various NLP tasks.
%Contrastive learning~\cite{1640964} further improves representation learning by pulling semantically similar examples together in the embedding space and separating dissimilar ones. We use this idea to fine-tune RoBERTa so that workflow contexts requiring the same reusable \textit{Action} obtain similar representations.
%While these pre-trained models provide rich contextual representations, their objectives mainly focus on token-level prediction rather than distinguishing semantically similar or dissimilar instances. 
%Contrastive learning~\cite{1640964} addresses this limitation 
Contrastive learning~\cite{1640964} further improves representation learning by introducing pairwise comparison, encouraging similar examples (positive pairs) to align closely in the embedding space while separating dissimilar ones (negative pairs). Given a batch of samples, an encoder transforms each input into a latent vector, and a similarity function (\eg cosine similarity) quantifies pairwise closeness. The resulting contrastive loss enhances semantic discrimination, which we leverage to fine-tune RoBERTa for more context-aware and task-specific representations in workflow step generation.

\subsection{LLMs and the ``Judge" Paradigm}

%In recent years, LLMs have emerged as transformative tools capable of understanding and generating human-like text from extensive pre-training data. Their versatility lies in the ability to generalize across diverse tasks using only textual instructions, without additional parameter updates. To further enhance their performance, researchers have developed \textit{prompt engineering} techniques~\cite{prompt_Ekin, liu2021pretrainpromptpredictsystematic} that strategically design input formulations to guide model behavior. The usage of LLMs can be differ by the level of task-specific context provided and can be broadly classified into the following three cases~\cite{brown2020languagemodelsfewshotlearners}: (i) \textbf{zero-shot} mode, where the model relies solely on pre-trained knowledge; (ii) \textbf{one-shot}, where a single example demonstrates the expected output is provided to the model; and (iii) \textbf{few-shot}, where multiple examples are given to refine reasoning and response style.

In recent years, LLMs have emerged as powerful tools capable of understanding and generating human-like text through extensive pre-training. Their versatility stems from the ability to generalize across diverse tasks using only textual instructions, without additional parameter updates. To further improve performance, researchers have proposed \textit{prompt engineering} techniques~\cite{prompt_Ekin, liu2021pretrainpromptpredictsystematic} that carefully design input prompts to guide model behavior. Depending on the amount of task-specific context provided, LLM usage can be categorized into three settings~\cite{brown2020languagemodelsfewshotlearners}: (i) \textbf{zero-shot}, relying solely on pre-trained knowledge; (ii) \textbf{one-shot}, where a single example is provided; and (iii) \textbf{few-shot}, where multiple examples are given to refine responses.
LLMs have transformed how developers interact with code, enabling natural-language-driven programming and automation. Models such as GPT-4~\cite{openai2024gpt4technicalreport}, Claude-Sonnet~\cite{claude_sonnet4}, CodeT5~\cite{wang-etal-2021-codet5}, and CodeBERT~\cite{feng2020codebertpretrainedmodelprogramming} demonstrate strong performance in code completion and synthesis. However, prior work largely focuses on general-purpose languages, leaving configuration artifacts underexplored. GHA workflows, written in YAML to define CI/CD automation, represent a distinct form of infrastructure configuration. Prior attempts to generate such workflows using fine-tuned T5 and LLMs~\cite{10.1145/3597503.3623351,10.1145/3664476.3664497} only partially capture workflow semantics, leading to limited accuracy and generalization. In this work, we guide LLMs to generate accurate and context-aware workflow steps for GHA automation.

Another emerging paradigm is \textit{LLM-as-a-Judge}, which uses large language models as automatic evaluators of generated content instead of relying solely on human annotators. This idea builds on the observation that instruction-tuned models can reason about and compare outputs in a human-like manner~\cite{zheng2023judgingllmasajudgemtbenchchatbot}. Recent surveys~\cite{gu2025surveyllmasajudge} describe LLM-as-a-Judge as a scalable and cost-effective alternative to manual evaluation, although challenges such as bias, inconsistency, and over-alignment remain. In software engineering, recent studies~\cite{ahmed2025llmsreplacemanualannotation,Wang_2025,zhou2025llmasjudgemetricbridginggap} explore its use for automatically evaluating code and software artifacts, suggesting that with proper calibration, LLMs can complement human evaluators in domain-specific tasks.

%%%%%%%%%%%%%%%%%%%%%%%
\section{\approach}
\label{sec:method}
%%%%%%%%%%%%%%%%%%%%%%%

In this study, we propose \textbf{\approach} (\textbf{F}ramework for \textbf{L}earning t\textbf{O} predict Workflow \textbf{STEP}s), a step-type-aware approach designed to complete GHA workflow steps. %a novel approach designed to complete GHA workflow steps. 
Given the preceding workflow context together with the descriptive name of the target step (if applicable), \approach recommends a workflow step that fulfills the intended functionality. %to recommend GHA workflow steps, combining a pre-trained RoBERTa model enhanced with contrastive learning and the state-of-the-art LLMs (\ie Claude-Sonnet). 

%In experiment settings, we represent each workflow as a sequence of steps extracted from its execution context, where each step is categorized as either an \textit{Action} or a \textit{Script}. Each instance is represented as a tuple \texttt{(input\_context, target\_step, type\_step)}, where \texttt{input\_context} denotes the preceding context. 

% \subsection{Overall Framework}

%\figref{fig:framework} presents the overall architecture of \approach. The framework combines two RoBERTa models with a state-of-the-art LLM to predict the next step in a workflow. First, a RoBERTa classifier determines the type of the next step, distinguishing between \textit{Action} and \textit{Script}. If the step is an \textit{Action}, a second RoBERTa model trained with contrastive learning recommends the most suitable \textit{Action}. If the step is a Script, an LLM generates the corresponding command required to fulfill the intended functionality. These components are integrated into a unified pipeline that predicts the next step given the workflow context. In the following subsections, we will describe each component of \approach in detail.

In our study, we represent each workflow as a sequence of steps extracted from its execution context. Specifically, each instance in our dataset \textit{D} is defined as a tuple \texttt{(input\_context, target\_step, type\_step)}, where \texttt{input\_context} denotes the preceding workflow context (\ie all workflow elements appearing before the target step concatenated with the descriptive step name), \texttt{target\_step} the next step to be generated, and \texttt{type\_step} its type (\textit{Action} or \textit{Script}). This formulation reflects the central modeling challenge of GHA step completion: a target step may require either selecting a reusable \textit{Action} identifier from the ecosystem or generating a project-specific shell \textit{Script}. We further derive two task-specific datasets from $D$: $D_{Action}$ and $D_{Script}$, which contain samples of type \textit{Action} and \textit{Script}, respectively.

%In our study, we represent each workflow as a sequence of steps extracted from its execution context. Specifically, each instance in our dataset \textit{D} is defined as a tuple \texttt{(input\_context, target\_step, type\_step)}, where \texttt{input\_context} denotes the preceding workflow context (\ie all workflow elements appearing before the target step concatenated with the descriptive step name), \texttt{target\_step} the next step to be generated, and \texttt{type\_step} its type (\textit{Action} or \textit{Script}). This formulation reflects the central modeling challenge of GHA step completion: the same workflow context may require either selecting a reusable \textit{Action} identifier from the ecosystem or generating a project-specific shell \textit{Script}. $D$ is further divided into two subsets: $D_{Action}$ and $D_{Script}$, which contain samples of type \textit{Action} and \textit{Script}, respectively.

\begin{figure}[h!]
 \centering
     % https://docs.google.com/drawings/d/1BVVn6N1aVF935fyLCrYouC2i6VEk6-vRG7j5qzollHY/edit?usp=sharing
     {\includegraphics[width=0.76\linewidth]{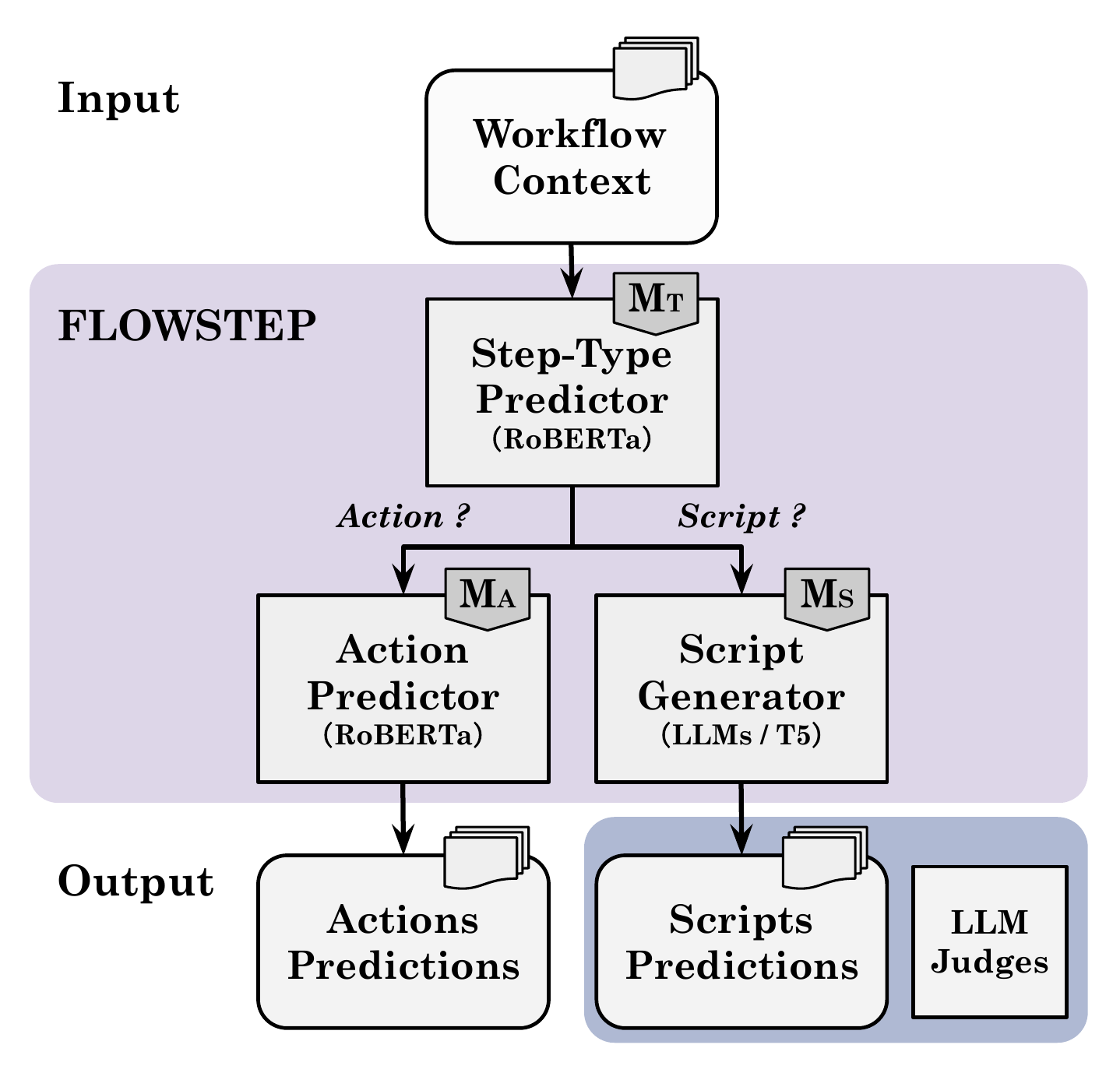}}
 \caption{Overall framework of \approach}
\label{fig:framework}
\end{figure}

Building on this formulation, \figref{fig:framework} presents the overall architecture of \approach. The framework integrates two RoBERTa models with a state-of-the-art LLM in a unified pipeline, so that each subtask is handled by a model aligned with its output space. Given an \texttt{input\_context}, a RoBERTa classifier first predicts the step type (\texttt{type\_step}). If the predicted type is \textit{Action}, a second RoBERTa model trained with contrastive learning recommends the most suitable \textit{Action}. Otherwise, an LLM generates the corresponding \textit{Script} to fulfill the intended functionality. Using the predicted scripts, we further investigate the feasibility of the LLM-as-a-Judge paradigm in our evaluation setting. In the following subsections, we describe each component of \approach in detail.

%Building on this formulation, \figref{fig:framework} presents the overall architecture of \approach. The framework integrates two RoBERTa models with a state-of-the-art LLM in a unified pipeline, so that each subtask is handled by a model aligned with its output space. Given an \texttt{input\_context}, a RoBERTa classifier first predicts the step type (\texttt{type\_step}). If the predicted type is \textit{Action}, a second RoBERTa model trained with contrastive learning recommends the most suitable \textit{Action}. Otherwise, an LLM generates the corresponding \textit{Script} to fulfill the intended functionality. Using the predicted scripts, we further investigate the feasibility of the LLM-as-a-Judge paradigm in our evaluation setting. In the following subsections, we describe each component of \approach in detail.

\subsection{Step-Type Predictor}
The design of \approach is motivated by an empirical analysis of the limitations in prior GHA recommendation models. Mastropaolo \etal~\cite{10.1145/3597503.3623351} analyzed 384 workflow steps incorrectly predicted by their step prediction model GH-WCOM, and found that 159 recommendations were genuinely incorrect, while the remaining errors involved only minor issues. In this study, we further inspected these 159 erroneous predictions and found that 33.3\% (53/159) are mismatches between \textit{Actions} and \textit{Scripts}, \ie cases where an expected \textit{Action} was replaced with a \textit{Script}, or vice versa. For example, the step was predicted to be a script execution ``\texttt{run: pip install tox}'' while the developers adopted an existing \textit{Action} ``\texttt{uses: actions/cache@v1}''.   % \listref{lst:mismatch} shows a real example where an \textit{Action} was wrongly predicted as a \textit{Script}.

% \begin{lstlisting}[
%   style=jsonstyle,
%   basicstyle=\ttfamily\small,
%   aboveskip=4pt,
%   belowskip=4pt,
%   lineskip=-1pt,
%   xleftmargin=0pt,
%   xrightmargin=0pt,
%   framexleftmargin=0pt,
%   framexrightmargin=0pt,
%   caption={Example of an expected Action has been wrongly predicted as a Script\footnotemark},
%   captionpos=b,
%   label={lst:mismatch},
%   frame=tb
% ]
% <HL2>Workflow context:</HL2>
% name: tests
% on: ...
% jobs: 
%   tests: ...
%   style:
%     runs-on: ubuntu-latest
%     steps: ...
%       <HL><TO_BE_PREDICTED></HL>
      
% <HL2>Action expected:</HL2> actions/cache@v1
% <HL2>Script generated:</HL2> pip install tox

% \end{lstlisting}
% \footnotetext{Due to space limitations, parts of the workflow context are omitted; please refer to our replication package for the full example.}

Based on this, we hypothesize that explicitly distinguishing between \textit{Action} and \textit{Script} steps can improve the recommendation accuracy by preventing a single model from conflating two different prediction spaces. Therefore, we implement a \textit{Step-Type Predictor} based on a RoBERTa encoder, mapping a workflow context (\ie \texttt{input\_context}) to a contextualized representation. This representation is passed through a dropout layer and a fully connected layer to perform binary classification of step type ($1$: \textit{Action}, $0$: \textit{Script}). The RoBERTa model is initialized with pre-trained weights from Hugging Face and fine-tuned using Binary Cross-Entropy with Logits Loss.

%Based on this observation, we hypothesize that explicitly distinguishing between \textit{Action} and \textit{Script} steps can improve overall recommendation accuracy. 

%To address this, we implement a \textit{Step-Type Predictor} based on a RoBERTa encoder. The model maps a workflow context (\ie \texttt{input}) to a representation $V_i$, followed by a dropout and fully connected layer for binary classification of \texttt{step\_type} ($1$ for \textit{Action}, $0$ for \textit{Script}). The classifier is initialized with pre-trained weights from Hugging Face and fine-tuned using the Binary Cross-Entropy with Logits Loss.

During FLOWSTEP inference, the \textit{Step-Type Predictor} acts as a router: it determines the step type for a given context and forwards the input to the corresponding downstream component: either the \textit{Action Predictor} or the \textit{Script Generator}.
% The trained classifier is subsequently employed to infer the step type from a given input workflow context, thereby determining the appropriate downstream model, namely the \textit{Action Predictor} or the \textit{Script Generator}, to which the context is forwarded.

\subsection{Action Predictor with Contrastive Learning}

The \textit{Action Predictor} in \approach consists of two primary components: (i) a RoBERTa-based \textbf{context encoder} that employs contrastive learning to align workflow contexts referring to the same \textit{Actions} while pushing apart those referring to distinct \textit{Actions}; and (ii) a \textbf{classification head} that maps these high-dimensional embeddings into specific \textit{Action} categories. During model development, the context encoder and classification head are trained in two distinct phases:

\textbf{Phase 1: Contrastive training of the context encoder.}
We fine-tune RoBERTa using a contrastive objective that combines N-pair Loss and Soft-Nearest Neighbor Loss~\cite{9796359}. For each \textit{Action} instance, we pair its workflow context with another context using the same \textit{Action} as a positive example, while contexts using different \textit{Actions} in the same batch serve as negative examples. This encourages contexts that require the same reusable \textit{Action} to be close in the embedding space and separates contexts associated with different \textit{Actions}; implementation details and the full loss formulation are provided in our replication package.  %This is a new sentence.

\textbf{Phase 2: Training the classification head.}
We freeze the encoder to preserve the learned semantic space, and attach a classification head consisting of a feed-forward network with ReLU activation and dropout ($p=0.2$). The model is then fine-tuned on $D_{Action}$ using Cross-Entropy loss to map context representations to \textit{Action} identifiers.
%In phase 2, we freeze the parameters of the contrastive encoder to preserve the learned semantic space. We attach a classification head, comprising a feed-forward network with a ReLU activation and a dropout layer ($p=0.2$), and train it on $D_{Action}$ again using standard Cross-Entropy loss. This allows the model to map the robust context embeddings to the final set of Action identifiers.

%Again, we initialize our model with the default parameters used in the Hugging Face's implementation of RoBERTa\footnote{https://huggingface.co/docs/transformers/model\_doc/roberta}, and fine-tune the model using contrastive loss.

\subsection{LLM-based Script Generator}

Unlike \textit{Actions}, \textit{Script} steps offer greater flexibility, making them more error-prone and harder to predict. Since modern LLMs are good at code generation and \textit{Scripts} essentially consist of shell commands, we leverage LLMs to generate executable workflow \textit{Scripts}.

\textbf{Model Selection.} We evaluate five LLMs spanning different providers and capability tiers with diverse training backgrounds: Claude-Sonnet-4.5, Claude-Haiku-4.5, DeepSeek-V3.2-Experimental, GPT-4.1, and GPT-3.5-turbo (see \tabref{tab:LLMs}). This selection covers both closed-weight (GPT, Claude) and open-weight (DeepSeek) models, as well as varying model scales. All models are accessed via their official APIs.

\begin{table}[t]
\centering
\small

\setlength\tabcolsep{3pt}
\renewcommand{\arraystretch}{1.1}
\caption{Large Language Models selected for our studies.}
\begin{tabular}{l|c|c|c}
\toprule
\textit{Model} & \textit{API Pricing*} &\textit{Released} & \textit{Provider}\\
\midrule
Claude-Sonnet-4.5 & \$3 / \$15  & Sept 2025 & Anthropic \\
Claude-Haiku-4.5 & \$1 / \$5  & Oct 2025 & Anthropic \\
DeepSeek-V3.2-Exp & \$0.28 / \$0.42  & Sept 2025 & DeepSeekAI \\
GPT-4.1 & \$2 / \$8  & Apr 2025 & OpenAI \\
GPT-3.5-turbo & \$0.5 / \$1.5  & Nov 2023 & OpenAI \\
\bottomrule
\end{tabular}
\\[3pt]
\parbox{\linewidth}{\footnotesize * Price for Input/Output per 1M tokens at experiment time}
\label{tab:LLMs}
\vspace{-15pt}
\end{table}

Each model is prompted in a tuning-free setting for \textit{Script} generation. The prompts differed between the raw and abstracted data: for the latter, five placeholder types used in the abstraction process were explicitly included. All prompts instructed the model to produce valid \textit{Script} steps formatted as shell commands, without explanations or commentary. To ensure fairness, all models were given identical input structures and generation constraints.

\textbf{Experimented prompts.} We use \textit{zero-shot} and \textit{one-shot} prompting to examine the effect of contextual examples on model performance. In the zero-shot setting, models rely solely on pre-trained knowledge, whereas the one-shot setting includes a single example demonstrating the desired input–output format. The full prompts are available in our replication package.

We focus on zero-shot and one-shot prompting to keep the comparison controlled across models and to avoid introducing additional prompt-selection factors from multiple demonstrations; broader few-shot or advanced prompting strategies are left for future work. Generation parameters were fixed across all models, with temperature set to 0 to reduce sampling variability. Each model generated one candidate \textit{Script} per instance, and outputs were post-processed for consistent formatting.

%Few-shot prompting is excluded, as it offered no clear gains over one-shot prompting while incurring substantially higher token costs. Generation parameters were fixed across all models, with temperature set to 0 for deterministic outputs. Each model generated one candidate \textit{Script} per instance, and outputs were post-processed for consistent formatting.

In addition to the five LLMs, a Text-to-Text Transfer Transformer (T5) model is used as a baseline and offline alternative, following the same architecture as GH-WCOM proposed in prior study~\cite{10.1145/3597503.3623351}. However, unlike GH-WCOM, which is fine-tuned on the full dataset $D$, % comprising both \textit{Action} and \textit{Script} samples, 
our T5 model is trained only on $D_{Script}$ to focus specifically on \textit{Script} generation, reducing interference from unrelated \textit{Action} data.

\textbf{Calibration set.} A small but representative calibration set $D_{Cali}$ of 1,000 randomly sampled test instances from $D_{Script}$ was used for comparative analysis to identify the most effective model and prompting configuration. Generated steps were compared against reference steps (\ie \texttt{target\_step}) using accuracy and n-gram based metrics (BLEU-4, ROUGE-L). The LLM with the best performance on $D_{Cali}$ is selected as the final Script Generator. \textit{Our results show that Claude-Sonnet-4.5 with one-shot prompting performs best; thus, it is used as the Script Generator in \approach.}%  for \approach.
% and used in the following experiments.

%Considering the inference cost of each LLM shown in~\tabref{tab:LLMs}, the model with the best price-to-performance ratio was selected as the final Script Generator for \approach, and will be used in the following experiments.

%All prompts, generation settings, and anonymized outputs are made publicly available to ensure reproducibility and facilitate future replication or extension studies.
%%%%%%%%%%%%%%%%%%%%%%%
\section{FLOWSTEP Evaluation}
\label{sec:design}
%%%%%%%%%%%%%%%%%%%%%%%

% We evaluate FLOWSTEP on the task of next-step prediction in GHA workflows using the test datasets summarized in~\tabref{tab:dataset}.

The \textit{goal} of this section is to evaluate the effectiveness of \approach in generating accurate and useful predictions. %Specifically, we assess its capability to generate the next step in a GHA workflow file, simulating the top-down coding process developers follow when constructing workflows. %The evaluation is carried out on the test datasets summarized in~\tabref{tab:dataset}.

\subsection{Research Questions}
% In order to understand workflow and job reruns in GHA, we formulated three research questions.
% We formulated the following research questions (RQs) to evaluate the effectiveness and architectural design of \approach:
We aim to answer the following research questions (RQs):

\textbf{RQ$_1$: \rqa}
This RQ examines whether the steps recommended by \approach align with those originally created by developers, using the previously proposed workflow completion model, GH-WCOM, as a baseline for comparison.

\textbf{RQ$_2$: \rqb}
%\textit{To what extent does explicitly distinguishing between Actions and Scripts improve the recommendation accuracy of FLOWSTEP?}
The architecture of \approach is based on a ``divide-and-conquer'' strategy that separates \textit{Action} and \textit{Script} generation. This RQ focuses on isolating the design choices of our framework through a series of ablation studies. We aim to quantify the contribution of each component to the overall performance of \approach.
%In this RQ, we isolate the core design choice of our framework to evaluate whether explicitly distinguishing between Actions and Scripts yields a measurable improvement in predictive performance.

%From the results of the previous RQs, we observe that improving raw accuracy in script generation is difficult and sometimes even misleading, as scripts may differ substantially in form while achieving the same functionality. This RQ investigates whether adopting the LLM-as-a-judge paradigm can provide a more meaningful evaluation by capturing semantic equivalence and practical utility beyond surface similarity.

\subsection{Data Preparation}

To ensure comparability with prior work, we use the dataset established by Mastropaolo \etal~\cite{10.1145/3597503.3623351} for GH-WCOM. The dataset contains 73,708 GHA workflows collected by Decan \etal~\cite{actiondata} and 733 additional workflows mined from 1,124 repositories by Mastropaolo \etal Their preprocessing removed duplicates, filtered workflows exceeding 1,024 tokens, and excluded non-ASCII instances, resulting in 17,935 unique workflows. Workflows are represented as JSON-like objects to preserve their structure.
From these workflows, we construct step-level instances for the prediction task. Each step-level instance is represented as a tuple \texttt{(input\_context, target\_step, type\_step)}, as mentioned in Section \ref{sec:method}. 

We split the dataset into training, validation, and test sets at the repository level using an 8:1:1 ratio, yielding 14,348/1,793/1,794 workflows and 108,899/13,009/13,630 step-level instances, respectively. 51.8\% of steps invoke \textit{Actions} and 48.2\% execute \textit{Scripts}. The broad dataset $D$ is used for step-type prediction, where only the binary \textit{Action}/\textit{Script} label is needed. $D_{Action}$ and $D_{Script}$ are task-specific datasets with additional filters, and their sizes are not expected to sum to $D$. 

For $D_{Action}$, we keep valid \texttt{owner/repo@version} \textit{Actions}, retain only labels appearing more than five times in the training set to focus on commonly reused \textit{Actions} with sufficient supervision for multi-class and contrastive learning, and deduplicate repeated contexts within each label. In the training split, this reduces 53,549 valid action uses to 51,251 covered contrastive instances across 332 labels. 

For $D_{Script}$, we keep steps with a non-empty \texttt{run} field, yielding 52,514 training instances. The number of instances in each dataset is shown in~\tabref{tab:dataset}.

%We then split the dataset into training, validation, and test sets using an 8:1:1 ratio. To prevent data leakage, we perform the split at the repository level, ensuring that all workflows from a single repository are confined to the same subset. This yields 14,348 training, 1,793 validation, and 1,794 testing workflows, corresponding to 108,899, 13,009, and 13,630 step-level instances, respectively. Overall, 51.8\% of steps invoke \textit{Actions} and 48.2\% execute \textit{Scripts}. For model training, the dataset $D$ is further divided into $D_{Action}$ and $D_{Script}$ based on step type. The number of instances in each dataset is shown in~\tabref{tab:dataset}.

Following prior work, we use both raw and abstracted data variants. The raw version preserves the original workflow content without modification. In the abstracted version, project-specific tokens such as paths, URLs, and SHAs are replaced with placeholders (e.g., \texttt{<PATH>}, \texttt{<URL>}), while preserving options, environment variables, and command syntax to improve generalization.

\begin{table}[h!]
\small
\centering
\renewcommand{\arraystretch}{1.1}
\caption{Number of instances in the used datasets}
\begin{tabular}{l|ccc}
\toprule
\textit{Dataset} & \textit{train} &\textit{validation} & \textit{test}\\
\midrule
$D$ & 108,899  & 13,009 & 13,630 \\
$D_{Action}$ & 51,251 & 5,984 & 6,256 \\
$D_{Script}$ & 52,514 & 6,297 & 6,532 \\
$D_{Cali}$ & - & - & 1,000 \\
\bottomrule
\end{tabular}
\label{tab:dataset}
\end{table}

\subsection{Evaluation Metrics}

To address RQ$_1$, we evaluate the best-performing configuration of \approach on the test sets. For each test instance, \approach generates the next workflow step based on the preceding workflow context, and the generated step is compared with the reference step from the original workflow. The overall performance is measured using exact-match accuracy, defined as the proportion of cases where the generated step exactly matches the reference step. To further analyze \approach, we evaluate its three sub-models separately on their respective prediction tasks.

For the \textit{Step-Type Predictor} ($M_T$), we report \textbf{accuracy}, defined as the proportion of correctly classified step types. The predicted step types are then used to route each instance to either \textit{Action Predictor} ($M_A$) or \textit{Script Generator} ($M_S$) for the subsequent step prediction.

%Misclassifications at this stage propagate to later tasks, thereby influencing the end-to-end performance of the overall approach.

To evaluate $M_A$, we again use \textbf{accuracy}. \textit{Action} instances misclassified as \textit{Script} by $M_T$ are counted as incorrect. For the remaining instances, a prediction is considered correct if the predicted action matches the ground-truth action.

For $M_S$, we also report \textbf{accuracy} of the generated scripts. \textit{Script} instances misclassified as \textit{Action} by $M_T$ are counted as incorrect predictions. For the remaining instances, we further evaluate generation quality using two widely adopted NLP metrics: \textbf{BLEU}~\cite{papineni-etal-2002-bleu} and \textbf{ROUGE}~\cite{lin-2004-rouge}. Following prior work in software engineering, we report BLEU-4, which measures 4-gram overlap between generated and reference scripts, and ROUGE-L, which evaluates similarity based on the longest common subsequence.

%For $M_S$, scripts misclassified as Actions by $M_T$ are counted as incorrect predictions. For the remaining cases, we report the \textbf{accuracy} of the generated scripts and also evaluate using two widely adopted NLP metrics: \textbf{BLEU}~\cite{papineni-etal-2002-bleu} and \textbf{ROUGE}~\cite{lin-2004-rouge}. Following prior work in software engineering, we report BLEU-4, which measures 4-gram overlap between generated and reference scripts, and ROUGE-L, which evaluates similarity based on the longest common subsequence.

\begin{comment}
For the Script generation model ($M_S$), as with Action prediction, scripts misclassified as Actions by $M_T$ are considered incorrect. For the remaining correctly identified cases, we assess the prediction quality of \approach using standard NLP evaluation metrics, namely BLEU~\cite{papineni-etal-2002-bleu} and ROUGE~\cite{lin-2004-rouge}.

\textbf{BLEU} (Bilingual Evaluation Understudy) measures the similarity between generated and reference texts by comparing overlapping $n$-grams, with scores ranging from 0 (no overlap) to 1 (identical). Following prior work in software engineering~\cite{}, we adopt the BLEU-4 variant.

\textbf{ROUGE} (Recall-Oriented Understudy for Gisting Evaluation) evaluates generated text quality by comparing it with reference outputs. We employ ROUGE-L, which measures the longest common subsequence between the generated and reference texts.

\end{comment}
To answer RQ$_2$, we investigate how different components and design choices affect the overall performance of \approach. We first examine the core design choice of explicitly distinguishing between \textit{Action} and \textit{Script} steps to determine whether this separation improves prediction performance. Specifically, we conduct an error analysis of the baseline model GH-WCOM to identify mispredictions associated with incorrect step-type classification. We then evaluate to what extent \approach reduces such errors by introducing a dedicated $M_T$ for step-type prediction.

To further understand the contribution of individual components in \approach, we conduct ablation studies on $M_A$ and $M_S$, while keeping $M_T$ fixed.

%To further understand how each component contributes to the overall performance of \approach, we conduct ablation studies on its three sub-models: Type Prediction ($M_{Type}$), Action Prediction ($M_{Action}$), and Script Generation ($M_{Script}$).

For $M_A$, we evaluate the impact of contrastive learning and the classification head on model performance by comparing the full model with two ablated variants:
\begin{itemize}
\item \textbf{No-classification head:} RoBERTa fine-tuned using contrastive learning without a classification head. \textit{Actions} are suggested based on cosine similarity between the input representation and the mean embedding of each \textit{Action} computed from the training set.
\item \textbf{No-contrastive:} RoBERTa fine-tuned without contrastive learning, where the model is trained using Mean Squared Error (MSE) loss to predict \textit{Action} representations.
\end{itemize}
We evaluate $M_A$ and its two variants on the test set of $D_{Action}$ using \textbf{Top-$K$ accuracy}~\cite{robillard2009recommendation}, which measures whether the ground-truth action appears among the top-$K$ recommendations ($K = 1$–$5$).

As for $M_S$, we evaluate the five selected LLMs (\tabref{tab:LLMs}) for \textit{Script} generation. We include a fine-tuned T5-small model trained on $D_{Script}$ as an offline alternative, and use the scripts generated by the original GH-WCOM as the baseline. All models are evaluated under identical setups on the same calibration test set $D_{Cali}$ using accuracy, BLEU, and ROUGE.

%As for $M_S$, we evaluate the impact of different state-of-the-art large language models (LLMs) on script generation quality using the 5 selected LLMs. As a controlled baseline, we train a T5-small model solely on $D_{Script}$. Additionally, we use the scripts generated by the original GH-WCOM. We then compare the models' performance under identical setups (same prompting for all LLMs and same input used for the T5-small model) on the same calibration test set, using raw accuracy, BLEU and ROUGE to assess how model scale and reasoning capability influence generation performance.

\subsection{Experimental Setting}

$M_T$ and $M_A$ are implemented in PyTorch based on the RoBERTa model provided by Hugging Face\footnote{\url{https://huggingface.co/}}. For all training phases of $M_T$ and $M_A$ (Phase 1 and Phase 2), the models were trained for 10 epochs with a batch size of 32. 
For the $M_S$ component, the temperature was set to 0 to ensure deterministic outputs across all five selected LLMs. The local T5-small model was trained for 70 epochs until convergence with a batch size of 8.
All trainable models were optimized using Adam with a learning rate of $2e^{-5}$. To ensure reproducibility, the random seed was fixed to 123456.

%$M_T$ and $M_A$ are implemented in PyTorch based on the RoBERTa model provided by Hugging Face\footnote{\url{https://huggingface.co/}}. For all training phases of $M_T$ and $M_A$ (Phase 1 and Phase 2), the models were trained for 10 epochs with a batch size of 32 using the Adam optimizer with a learning rate of $2e^{-5}$. To ensure reproducibility, the random seed was fixed to 123456.
%For the $M_S$ component, we set the temperature to 0 to ensure deterministic outputs across all five selected LLMs. The local T5-small model was trained for 70 epochs until convergence, using a batch size of 8 and the Adam optimizer with a learning rate of $2e^{-5}$.

%%%%%%%%%%%%%%%%%%%%%%%
\subsection{Experiment Results}
\label{sec:results}

% This section presents the results for RQ1 and RQ2.

%This section presents the experimental results for RQ1 and RQ2, including the overall effectiveness of \approach, comparison with GH-WCOM, the impact of distinguishing \textit{Action} and \textit{Script} steps, and the ablation study.

%This section presents the experimental results addressing the first two RQs. We evaluate the overall effectiveness of \approach, compare it with the baseline model GH-WCOM, analyze the impact of explicitly distinguishing between Action and Script steps, and report the results of the ablation study.

\subsubsection{RQ1: \rqa}

\begin{table*}[t]
\small
\centering
\renewcommand{\arraystretch}{1.1}
\caption{Overall performance comparison between FLOWSTEP and GH-WCOM on workflow step prediction.}
\label{tab:rq1_overall}
\begin{tabular}{c|l|rrrr|rr}
\toprule
\textit{Dataset} & \textit{Model} & \textit{Overall} & $M_{T}$ & $M_{A}$ & $M_{S}$ & \textit{BLEU} ($M_{S}$) & \textit{ROUGE} ($M_{S}$)\\
\midrule
\multirow{3}*{Abs} & $\textbf{FLOWSTEP}_{Claude-Sonnet-4.5}$ & \textbf{51.05\%} & 92.9\% & 76.28\% & 24.72\% & \textbf{35.36\%} & \textbf{59.26\%} \\
~ & FLOWSTEP$_{T5}$ & 48.0\% & 92.9\% & 76.28\% & 18.49\% & 11.6\% & 45.05\% \\
~ & GH-WCOM & 39.33\% & 82.86\% & 61.71\% & 15.0\% & 14.83\% & 38.46\% \\
\midrule
\multirow{3}*{Raw} & $\textbf{FLOWSTEP}_{Claude-Sonnet-4.5}$ & \textbf{51.09\%} & 91.96\% & 76.58\% & 24.48\% & \textbf{40.82\%} & \textbf{60.81\%} \\
~ & FLOWSTEP$_{T5}$ & 46.66\% & 91.96\% & 76.58\% & 15.42\% & 18.49\% & 42.26\% \\
~ & GH-WCOM & 38.23\% & 82.28\% & 61.09\% & 13.4\% & 13.14\% & 37.33\%\\
\bottomrule
\end{tabular}
\end{table*}

Table~\ref{tab:rq1_overall} summarizes the overall performance of \approach as an end-to-end pipeline, where \textit{Overall} presents the accuracy of records across both types, $M_{T}$, $M_{A}$ and $M_{S}$ stand for the accuracy separately for \textit{Type}, \textit{Action} and \textit{Script} predictions. For abstracted data, FLOWSTEP achieves an overall accuracy of 51.05\%, substantially exceeding GH-WCOM (39.33\%). %When disaggregated by step type, FLOWSTEP achieves 76.28\% for Actions and 18.16\% for Scripts, outperforming GH-WCOM’s 61.71\% and 15.0\%, respectively.
A similar trend is observed for raw data, where FLOWSTEP attains an overall accuracy of 51.09\%, compared to 38.23\% for the baseline. %Specifically, FLOWSTEP reaches 76.58\% for Actions and 13.52\% for Scripts, surpassing GH-WCOM’s 61.09\% and 13.4\%.
To disentangle where improvements arise, we further evaluate each of FLOWSTEP’s components in isolation:

\begin{itemize}
    \item \textbf{Step-Type Predictor ($M_{T}$):} The RoBERTa-based classifier effectively distinguishes between \textit{Action} and \textit{Script} steps, achieving 92.9\% accuracy on the abstracted dataset and 91.96\% on the raw dataset, substantially outperforming GH-WCOM (82.86\% and 82.28\%, respectively). This strong classification performance provides a reliable foundation for the subsequent modules of \approach.
    \item \textbf{Action Predictor ($M_{A}$): } For \textit{Action} prediction, the contrastive learning–enhanced predictor substantially improves retrieval accuracy over GH-WCOM, achieving Top-1 accuracies of 76.28\% (abstracted) and 76.58\% (raw), compared to 61.71\% and 61.09\% achieved by the baseline.
    \item \textbf{Script Generator ($M_{S}$):} For \textit{Script} generation alone, \approach (with LLM) achieves accuracies of 24.72\% (abstracted) and 24.48\% (raw), compared to 15.0\% and 13.4\% obtained by the baseline GH-WCOM. The offline T5-small model achieves 18.49\% and 15.42\%, outperforming GH-WCOM but still falling short of the LLM-based approach. Despite this improvement, the overall accuracy is still relatively low, suggesting that script generation remains challenging.

    %For script generation alone, \approach achieves accuracies of 24.72\% (abstracted) and 24.48\% (raw), compared to 15.0\% and 13.4\% obtained by GH-WCOM. These results suggest that LLM model produces more accurate scripts, but the performance gain remains limited. %This demonstrates that GPT-4.1 is able to generate more accurate scripts, but the improvement is not substantial. 
\end{itemize}

Taken together, by reducing noise from irrelevant data and leveraging specialized submodels, \approach achieves consistently superior performance across all metrics for GHA workflow step recommendation compared to GH-WCOM.

\subsubsection{RQ2: \rqb}

%Our approach contains three sub-models $M_{Type}$, $M_{Action}$ and $M_{Script}$. 
In this RQ, we investigate the impact of different components and design choices within the three models ($M_{T}$, $M_{A}$, and $M_{S}$) on the overall performance of \approach. Specifically, RQ$_2$ is divided into the following three sub-questions:

\begin{itemize}
    \item \textbf{RQ$_{2.1}$}: To what extent does $M_{T}$ reduce step-type confusion errors in workflow step prediction?
    
    %To what extent does explicitly distinguishing between \textit{Action} and \textit{Script} steps improve the overall performance of \approach?
    \item \textbf{RQ$_{2.2}$}: What is the contribution of contrastive learning and the classification head to the performance of $M_{A}$?
    \item \textbf{RQ$_{2.3}$}: How do different LLMs and prompting strategies (zero-shot vs. one-shot) affect the performance of $M_{S}$?
\end{itemize}

\paragraph{Impact of Step Type Distinction (RQ$_{2.1}$)}

\begin{comment}
\begin{table*}[t]
%\setlength\tabcolsep{4pt}
\renewcommand{\arraystretch}{1.1}
\centering
\caption{Error breakdown of GH-WCOM vs. FLOWSTEP on Step Type (ST) mismatch.}
\label{tab:rq2_errors}
\begin{tabular}{l|c|cc}
\toprule
\textit{Error Type} & \textit{Dataset} & \textit{GH-WCOM} & \textit{FLOWSTEP} \\
\midrule
ST mismatch (Action $\rightarrow$ Script)  & \multirow{3}*{Abs} & 1105 (13.36\%) & 603 (8.66\%) \\
ST mismatch (Script $\rightarrow$ Action) & ~ & 1231 (14.89\%) & 345 (4.95\%) \\
ST mismatch (All) & ~ & 2336 (28.25\%) & 948 (13.61\%) \\
\midrule
ST mismatch (Action $\rightarrow$ Script)  & \multirow{3}*{Raw} & 1143 (13.58\%) & 644 (9.01\%) \\
ST mismatch (Script $\rightarrow$ Action) & ~ & 1272 (15.11\%) & 430 (6.02\%) \\
ST mismatch (All) & ~ & 2415 (28.69\%) & 1074 (15.03\%) \\
\bottomrule
\end{tabular}
\end{table*}
\end{comment}

\begin{table}[t]
\renewcommand{\arraystretch}{1.1}
\small
\centering
\caption{Error breakdown on Step-type mismatch.}
\label{tab:rq2_errors}
\begin{tabular}{c|l|rr}
\toprule
\textit{Dataset} & \textit{Error Type} & \textit{GH-WCOM} & \textit{FLOWSTEP} \\
\midrule
\multirow{3}*{Abs} & A $\rightarrow$ S & 1105 (13.36\%) & 603 (9.23\%) \\
~ & S $\rightarrow$ A & 1231 (14.89\%) & 345 (5.28\%) \\
~ & All & 2336 (28.25\%) & \textbf{948 (14.51\%)} \\
\midrule
\multirow{3}*{Raw} & A $\rightarrow$ S & 1143 (13.58\%) & 644 (9.86\%) \\
~ & S $\rightarrow$ A & 1272 (15.11\%) & 430 (6.58\%) \\
~ & All & 2415 (28.69\%) & \textbf{1074 (16.45\%)} \\
\bottomrule
\end{tabular}
\\[3pt]
\parbox{\linewidth}{\footnotesize * A $\rightarrow$ S: Action incorrectly predicted as Script. Percentages are computed over each model's total number of incorrect end-to-end predictions, not over the full test set.}
\end{table}

Unlike GH-WCOM, which processes all inputs uniformly, \approach first predicts the step type using $M_{T}$ and then invokes the corresponding model (\ie $M_{A}$ or $M_{S}$). This design aims to reduce confusion between \textit{Action} and \textit{Script} steps. To evaluate its effectiveness, we analyze and compare the step-type confusion errors made by \approach and GH-WCOM.

As shown in \tabref{tab:rq2_errors}, %the percentages report the share of each model's incorrect end-to-end predictions caused by step-type mismatches, rather than the size of the evaluation set. 
over 28\% of GH-WCOM's errors are due to step-type mismatches, with 2,336 (28.25\%) and 2,415 (28.69\%) cases in the abstracted and raw datasets, respectively. In particular, GH-WCOM misclassifies 1,105 (abstracted) and 1,143 (raw) \textit{Actions} as \textit{Scripts}, while misidentifying 1,231 (abstracted) and 1,272 (raw) \textit{Scripts} as \textit{Actions}. In contrast, \approach, equipped with an explicit Step-Type Predictor $M_{T}$, substantially reduces these errors by nearly half. 
On the abstracted dataset, it records only 603 \textit{Action}$\rightarrow$\textit{Script} mismatches (9.23\%) and 345 \textit{Script}$\rightarrow$\textit{Action} mismatches (5.28\%), totaling 948 errors (14.51\%) using the final 6,534 incorrect predictions as the denominator. On the raw dataset, these numbers are 644 \textit{Action}$\rightarrow$\textit{Script} mismatches (9.86\%) and 430 \textit{Script}$\rightarrow$\textit{Action} mismatches (6.58\%), totaling 1,074 errors (16.45\%) using the final 6,530 incorrect predictions as the denominator.

For completeness, these total mismatch errors correspond to 7.1\% and 8.04\% of all evaluated test instances for \approach in the abstracted and raw settings, respectively, whereas the corresponding rates for GH-WCOM are 17.14\% and 17.72\%. The markedly lower mismatch rates achieved by \approach demonstrate the effectiveness of explicitly distinguishing between \textit{Actions} and \textit{Scripts} before generating workflow steps.

%As shown in \tabref{tab:rq2_errors}, over 28\% of GH-WCOM’s prediction errors are due to step type mismatches, with 2,336 (28.25\%) and 2,415 (28.69\%) cases in the abstracted and the raw dataset, respectively. In particular, GH-WCOM misclassifies 1,105 (abstracted) and 1,143 (raw) \textit{Actions} as \textit{Scripts}, while misidentifying 1,231 (abstracted) and 1,272 (raw) \textit{Scripts} as \textit{Actions}. In contrast, \approach, equipped with an explicit Step-Type Predictor $M_{T}$, substantially reduces these errors by nearly half. 
%On the abstracted dataset, it records only 603 \textit{Action}$\rightarrow$\textit{Script} mismatches (8.66\%) and 345 \textit{Script}$\rightarrow$\textit{Action} mismatches (4.95\%), totaling 948 errors (13.61\%). On the raw dataset, these numbers are 644 \textit{Action}$\rightarrow$\textit{Script} mismatches (9.01\%) and 430 \textit{Script}$\rightarrow$\textit{Action} mismatches (6.02\%), totaling 1,074 errors (15.03\%).  

%As \approach also produces fewer overall prediction errors, these results highlight the clear advantage of explicitly distinguishing between \textit{Actions} and \textit{Scripts} before generating workflow steps.

\paragraph{Contribution of Contrastive Learning \& Classification Head to $M_{A}$ (RQ$_{2.2}$)}

\begin{table*} \centering
\small
\renewcommand{\arraystretch}{1.1}
\caption{Comparison of different variants of our approach for Action prediction using Top-k accuracy (k = [1–5])}
\begin{tabular}{ c|l|rrrrr } 
 \toprule
 \textit{Dataset} & \textit{Model} & \textit{k = 1} & \textit{k = 2} & \textit{k = 3} & \textit{k = 4} & \textit{k = 5} \rule{0pt}{2.2ex}\\ 
 \midrule
 \multirow{4}*{Abs} & Finetune w/\phantom{o}  Contrastive Learning + Classification head & \textbf{87.23\%} & \textbf{91.56\%} & \textbf{93.14\%} & \textbf{93.88\%} & \textbf{94.63\%} \\
 ~ & Finetune w/\phantom{o} Contrastive Learning & 80.05\% & 84.91\% & 86.94\% & 88.11\% & 89.11\%\\ 
 ~ & Finetune w/o Contrastive Learning & 21.51\% & 33.42\% & 41.78\% & 47.43\% & 50.3\%\\
 ~ & GH-WCOM & 61.71\% & - & - & - & -\\ 
 \midrule
 \multirow{4}*{Raw} & Finetune w/\phantom{o}  Contrastive Learning + Classification head & \textbf{88.08\%} & \textbf{91.27\%} & \textbf{92.77\%} & \textbf{93.69\%} & \textbf{94.17\%} \\
 ~ & Finetune w/\phantom{o}  Contrastive Learning & 83.57\% & 87.82\% & 89.47\% & 90.46\% & 91.26\%\\
 ~ & Finetune w/o Contrastive Learning & 23.37\% & 35.07\% & 43.05\% & 49.63\% & 52.49\%\\ 
 ~ & GH-WCOM & 61.09\% & - & - & - & -\\
 \bottomrule
\end{tabular}
\label{tab:action_solo}
\end{table*}

We conduct an ablation study comparing $M_A$ with its two variants (\textit{No-classification head} and \textit{No-contrastive}), evaluated using Top-$K$ accuracy on $D_{Action}$. \tabref{tab:action_solo} reports the results of the $M_{A}$ variants on both abstracted and raw datasets, using GH-WCOM as the baseline. Without contrastive learning, the model yields poor performance (Top-1: 21.51\%/23.37\%, Top-5: 50.30\%/52.49\%), showing that standard fine-tuning alone is insufficient for accurate \textit{Action} prediction. Introducing contrastive learning leads to substantial improvements, increasing Top-1 performance to 80.05\%/83.57\% and Top-5 to 89.11\%/91.26\%. Adding the classification head further improves Top-1 by 7.18\%/4.51\% and Top-5 by 5.52\%/2.91\%, with diminishing gains as $K$ increases. This trend is consistent across both abstracted and raw datasets, and the full model substantially outperforms GH-WCOM. Overall, contrastive learning accounts for most of the performance gain, while the classification head provides additional improvements, particularly for Top-1 accuracy. The high Top-5 accuracy of the full model (94.63\%/94.17\%) further shows that $M_A$ can provide a compact ranked set of plausible reusable \textit{Actions}, which is useful when the top-ranked prediction is not correct but the correct \textit{Action} remains among the leading candidates.

\paragraph{Impact of Model Choice on $M_{S}$ (RQ$_{2.3}$)}

\begin{table}[t]
\centering
\small
\setlength{\tabcolsep}{4pt}
\renewcommand{\arraystretch}{1.1}
\caption{Comparison of model choices for Script generation on the calibration set under zero-shot prompting}
\begin{tabular}{c|l|rrr}
\toprule
\textit{Dataset} & \textit{Model} & \textit{Acc} & \textit{BLEU} & \textit{ROUGE} \\
\midrule
\multirow{7}{*}{Abs}
& Claude-Sonnet-4.5   & \textbf{24.4\%} & \textbf{31.03\%} & \textbf{57.68\%} \\
& Claude-Haiku-4.5    & 20.9\% & 29.4\%  & 53.66\% \\
& DeepSeek-V3.2-Exp   & 19.0\% & 29.23\% & 53.4\%  \\
& GPT-4.1             & 19.6\% & 23.7\%  & 53.62\% \\
& GPT-3.5-turbo       & 7.2\%  & 13.0\%  & 34.56\% \\
\cline{2-5}
& T5-small            & 18.8\% & 11.1\%  & 46.47\% \rule{0pt}{2.2ex}\\
& GH-WCOM             & 14.6\% & 11.22\% & 37.03\% \\
\midrule
\multirow{7}{*}{Raw}
& Claude-Sonnet-4.5   & \textbf{27.4\%} & 36.86\% & \textbf{60.54\%} \\
& Claude-Haiku-4.5    & 23.0\% & \textbf{39.63\%} & 58.5\%  \\
& DeepSeek-V3.2-Exp   & 19.2\% & 35.98\% & 54.95\% \\
& GPT-4.1             & 20.4\% & 31.56\% & 56.52\% \\
& GPT-3.5-turbo       & 11.5\% & 19.36\% & 41.31\% \\
\cline{2-5}
& T5-small            & 14.7\% & 18.17\% & 43.15\% \rule{0pt}{2.2ex}\\
& GH-WCOM             & 12.9\% & 11.73\% & 36.78\% \\
\bottomrule
\end{tabular}
\label{tab:zero_shot}
\end{table}

\begin{table}[t]
\centering
\small
\setlength{\tabcolsep}{4pt}
\renewcommand{\arraystretch}{1.1}
\caption{Comparison of model choices for Script generation on the calibration set under one-shot prompting}
\begin{tabular}{c|l|rrr}
\toprule
\textit{Dataset} & \textit{Model} & \textit{Acc} & \textit{BLEU} & \textit{ROUGE} \\
\midrule
\multirow{7}{*}{Abs}
& Claude-Sonnet-4.5   & \textbf{27.5\%} & \textbf{37.79\%} & \textbf{61.57\%} \\
& Claude-Haiku-4.5    & 21.3\% & 29.31\% & 54.15\% \\
& DeepSeek-V3.2-Exp   & 21.9\% & 32.08\% & 55.61\% \\
& GPT-4.1             & 23.4\% & 27.57\% & 55.6\%  \\
& GPT-3.5-turbo       & 5.2\%  & 10.04\% & 23.71\% \\
\cline{2-5}
& T5-small            & 18.8\% & 11.1\%  & 46.47\% \rule{0pt}{2.2ex}\\
& GH-WCOM             & 14.6\% & 11.22\% & 37.03\% \\
\midrule
\multirow{7}{*}{Raw}
& Claude-Sonnet-4.5   & \textbf{29.6\%} & \textbf{42.81\%} & \textbf{63.44\%} \\
& Claude-Haiku-4.5    & 23.9\% & 39.05\% & 59.68\% \\
& DeepSeek-V3.2-Exp   & 22.1\% & 38.61\% & 57.93\% \\
& GPT-4.1             & 24.7\% & 36.68\% & 59.12\% \\
& GPT-3.5-turbo       & 11.0\% & 18.4\%  & 40.64\% \\
\cline{2-5}
& T5-small            & 14.7\% & 18.17\% & 43.15\% \rule{0pt}{2.2ex}\\
& GH-WCOM             & 12.9\% & 11.73\% & 36.78\% \\
\bottomrule
\end{tabular}
\label{tab:one_shot}
\end{table}

\tabref{tab:zero_shot} and \tabref{tab:one_shot} present the performance of different models for Script generation on $D_{Cali}$ using both abstracted and raw data, evaluated with accuracy (Acc), BLEU, and ROUGE metrics, with GH-WCOM serving as the baseline. Under zero-shot prompting, Claude-Sonnet-4.5 achieves the best overall performance across most metrics, reaching 24.4\% and 27.4\% accuracy on abstracted and raw data, respectively, together with strong BLEU (31.03\%/36.86\%) and ROUGE (57.68\%/60.54\%) scores. Claude-Haiku-4.5 and DeepSeek-V3.2-Exp also perform competitively, while GPT-4.1 achieves comparable accuracy but lower BLEU on abstracted data. In contrast, GPT-3.5-turbo performs notably worse across all metrics, with accuracy (7.2\%/11.5\%) that remains below or only slightly above the baseline.

One-shot prompting leads to improvements for several models, although the effect is not uniform across all systems. In particular, Claude-Sonnet-4.5 further increases its accuracy to 27.5\% and 29.6\% on abstracted and raw data, respectively, while also achieving the highest BLEU (37.79\%/42.81\%) and ROUGE (61.57\%/63.44\%) scores among all models. DeepSeek-V3.2-Exp and GPT-4.1 also show moderate gains in both accuracy and lexical similarity metrics. However, the impact of one-shot prompting is less consistent for smaller or weaker models, such as GPT-3.5-turbo, where performance does not improve and in some cases slightly decreases. Overall, these results indicate that providing a single demonstration example can help guide stronger models toward better \textit{Script} generation, although the benefit depends on the model capacity.

%The T5-small model, trained solely for the \textit{Script} generation task, while behind most LLMs except GPT-3.5-turbo, performs competitively (18.8\%/14.7\% accuracy) and serves as a practical offline alternative that does not require API access, making it a suitable backup model for $M_{S}$.

The T5-small model, trained solely for the \textit{Script} generation task, while behind most LLMs except GPT-3.5-turbo, performs competitively (18.8\%/14.7\% accuracy) and serves as a practical offline alternative that does not require API access, making it a suitable backup model for $M_{S}$. Together with the API prices in \tabref{tab:LLMs}, these results illustrate a quality--deployment-cost tradeoff: Claude-Sonnet-4.5 offers the best generation quality, whereas T5-small serves as a lightweight local baseline. Open-weight LLMs also provide a viable self-hosted alternative and will be investigated in future work.

%whereas T5-small provides a lower-dependency option when external API calls are undesirable.} \BIN{you can also avoid this with open-weight models...}

%%%%%%%%%%%%%%%%%%%%%%%
\section{LLM-as-a-Judge}
\label{sec:judge}

RQ2.3 shows that even the best-performing LLM, Claude-Sonnet-4.5, achieves only around 29.6\% accuracy on $D_{Cali}$ for \textit{Script} generation. In practice, improving accuracy on this task is inherently difficult, and accuracy alone can be misleading for model performance. For example, in one case of our study, ``\texttt{pip install mythx-cli}'' is considered wrong when compared with the oracle ``\texttt{pip3 install mythx-cli}'' given the metrics used, while in reality both scripts may be functionally equivalent in some environments. 
% \listref{lst:mythx-workflow} shows a real example where the predicted \textit{Script} is labeled as incorrect by accuracy, despite being functionally equivalent to the expected outcome. 
%In general, \textit{Scripts} may differ substantially in form while achieving the same functionality. Therefore, in this section, we investigate whether the LLM-as-a-Judge paradigm can provide a more effective evaluation of generated \textit{Scripts} than commonly used metrics.

In general, \textit{Scripts} may differ substantially in form while achieving the same functionality. Thus, exact-match accuracy should be interpreted as a conservative lower bound rather than a complete measure of functional correctness. Therefore, in this section, we investigate whether the LLM-as-a-Judge paradigm can provide a more effective evaluation of generated \textit{Scripts} than commonly used metrics.

\begin{comment}
\begin{lstlisting}[
  style=jsonstyle,
  basicstyle=\ttfamily\small,
  aboveskip=4pt,
  belowskip=4pt,
  lineskip=-1pt,
  xleftmargin=0pt,
  xrightmargin=0pt,
  framexleftmargin=0pt,
  framexrightmargin=0pt,
  caption={Example of a ``wrongly'' predicted GHA workflow script that is functionally correct},
  captionpos=b,
  label={lst:mythx-workflow},
  frame=tb
]
name: Mythx
on: 
  workflow_dispatch: null
jobs: 
  mythx: 
    name: Submit to Mythx
    runs-on: ubuntu-latest
    steps: 
      ...
      name: Install mythx CLI
      <HL><TO_BE_PREDICTED></HL>
      
<HL2>Script expected:</HL2> pip3 install mythx-cli
<HL2>Script generated:</HL2> pip install mythx-cli

\end{lstlisting}
\end{comment}

\subsection{Study Design}

% The goal of this section is to address RQ$_3$: \textit{Can we employ LLM-as-a-judge to more effectively evaluate the steps generated by FLOWSTEP and other generative models?}

\subsubsection{Research Question}
To understand whether LLMs can serve as a reliable judge to evaluate the correctness of generated step scripts, we formulate the following research question:

\textbf{RQ$_3$: Can LLM-as-a-Judge more effectively evaluate automatically generated Script steps?}

%Can we employ LLM-as-a-judge to more effectively evaluate the steps generated by FLOWSTEP? and other generative models?}
% Results from the previous RQs show that raw accuracy is difficult to improve and can be misleading in script generation, as scripts may differ in form while achieving the same functionality. This RQ therefore investigates whether LLM-as-a-Judge provides a more meaningful evaluation by capturing semantic equivalence and practical utility beyond surface similarity.
%RQ$_3$: Can we employ LLM-as-a-judge to more effectively evaluate the steps generated by FLOWSTEP and other generative models?

\subsubsection{Model Selection}
The same five LLMs (\ie Claude-Sonnet-4.5, Claude-Haiku-4.5, DeepSeek-V3.2-Exp, GPT-4.1 and GPT-3.5-turbo) that we selected for \textit{Script} generation are tested as judges.
%\textbf{Claude-Sonnet-4.5 and Claude-Haiku-4.5}.

%\textbf{DeepSeek-V3.2-Exp}. DeepSeek-V3.2-Exp is part of the DeepSeek series of llms and has been shown to perform well on both natural language and code-related tasks. Being an experimental successor of DeepSeek-V3, the model has been further optimized with a sparse-attention mechanism to improve efficiency and support longer contexts. While detailed technical specifications are not fully disclosed, it is reasonable to assume that DeepSeek-V3.2-Exp remains comparable in scale to the V3 model family—reportedly trained on trillions of tokens comprising code, English, and Chinese text. The instruct versions of the model have been fine-tuned to better follow human instructions across dialogue and programming domains. We access the model through the DeepSeek API.

%\textbf{GPT-3.5-turbo and GPT-4.1}. Both models are behind OpenAI ChatGPT and have been shown to be able to handle code-related tasks [14], [26], [34]. Being the models on top of which a chatbot has been built, both GPT variants have been trained to follow instructions for required tasks. While the technical details behind these models are not publicly available, we can safely claim that (i) they are substantially larger than the DeepSeek Coder/CodeLlama variants previously described, with rumors placing GPT3.5 and GPT-4 at over 150B and 1.5T trainable parameters, respectively; and (ii) their training set features massive amounts of textual data, including code mined from GitHub repositories. We exploit these models through the ChatGPT APIs.

\subsubsection{Evaluation Datasets}
To assess the effectiveness of LLMs-as-a-Judge for evaluating \textit{Script} generation, we randomly sample 350 cases of incorrectly generated \textit{Scripts} produced by the five evaluated LLMs in \approach for raw data, with no overlap between models. The number of samples drawn from each model is proportional to the number of incorrect \textit{Scripts} produced by that model, ensuring a representative distribution of error cases. The sample size of 350 corresponds to a statistically significant subset, ensuring a 95\% confidence level with a confidence interval of $\pm5\%$. Each record consists of (i) the original input context that is fed to the generation model; (ii) a target step representing the original correct implementation; (iii) a generated step that is produced by LLMs; and (iv) two independent judgments provided by two of the authors. The records were classified using one of the following labels:

\begin{itemize}
\item \textbf{Functionally Correct (2): } The generated step performs the same purpose or achieves the same result as the ground-truth step.
\item \textbf{Functionally Almost Correct (1): } The generated step is mostly correct but requires a minor modification to be fully usable or achieve the intended functionality.
\item \textbf{Functionally Incorrect (0): } The generated step differs in purpose, behavior, or misses key functionality.
\end{itemize}

A comprehensive annotation guideline provided to the human annotators is available in our replication package. After finalizing the dataset, we reassessed inter-annotator agreement using Krippendorff’s $\alpha$, which is interpreted as follows:
\begin{itemize}
    \item $< 0.10:$ agreement equivalent to chance; 
    \item $0.10$–$0.20:$ weak agreement; 
    \item $0.21$–$0.40:$ fair agreement; 
    \item $0.41$–$0.60:$ moderate agreement; 
    \item $0.61$–$0.80:$ substantial agreement; 
    \item $0.81$–$0.99:$ near-perfect agreement; 
    \item $1 =$ perfect agreement.
\end{itemize}

For the functional correctness of the generated steps, the two primary annotators achieved $\alpha = 0.82$, indicating near-perfect agreement. To obtain reliable ground-truth labels for evaluating the LLM judges, disagreements were first resolved through discussion between the two annotators, for the remaining 18 cases (out of 350) where consensus could not be reached, a third author acted as adjudicator to assign the final labels.

%\textcolor{red}{For the functional correctness of the generated Scripts, the two authors obtained $\alpha = 0.82$}, indicating quite a high degree of agreement among the two authors. To obtain a ground truth for the evaluation of the LLMs judges, we carefully resolved the disagreements by discussion, and asking the third author to join as adjudicator.

\subsubsection{Data Collection}
%As for RQ$_3$, we explore the feasibility of adopting the LLM-as-a-judge paradigm to evaluate Script generation results more effectively. Instead of relying solely on surface-level metrics such as BLEU or ROUGE, we prompt advanced LLMs to assess the generated scripts in terms of their semantic equivalence, functional correctness, and practical utility compared to the reference implementations.

Using the manually labeled dataset, we then ask the LLMs to provide their judgments.
% \textbf{Experimented prompts.}
%To ensure that LLM judgments align with human assessments, we design structured evaluation prompts that provide the workflow context together with the ground-truth and generated steps (scripts), following the same guidelines used for human annotation. Consistent with the setup in RQ2, we evaluate both \textit{zero-shot} and \textit{one-shot} prompting.
%Based on previous studies, variety has been observed in the output template used by the LLMs, with some beings more verbose than others and providing additional unrequested information. To handle this, we limit the output of the LLMs by explicitly prompted them to output only a decision number. To be confident about the judgments given by the LLMs, all of them have been manually checked by two authors, Making sure all the judgements are actually valid.
To align LLM judgments with human assessments, we design structured prompts that present the workflow context along with the ground-truth and generated steps, following the same guidelines used for human annotation. As in RQ2, we evaluate \textit{zero-shot} and \textit{one-shot} prompting, with both prompts available in our replication package. As observed in prior studies~\cite{crupi2025effectivenessllmasajudgecodegeneration}, LLMs may produce verbose or inconsistent output formats, so we explicitly instruct them to return only a decision number (\ie 0, 1 or 2). All judgments were then manually verified by the first author to ensure their validity.

\begin{comment}
An example of a \textit{zero-shot} prompt is provided below:
\textit{You are an expert DevOps Engineer, specialized in the GitHub Action platform.
        Your task is to evaluate how well a generated GitHub Actions workflow step (called the "Output") matches the actual ground-truth step (called the "Target"), given the preceding workflow context ("Input"). The generated step is always a script.\\
        Carefully compare the output and the target, and choose only one of the following three resolution labels that best describe their functional similarity:\\
        (2) Functionally Correct: The generated step performs the same purpose or achieves the same result as the ground-truth step.\\
        (1) Functionally Almost Correct: The generated step is mostly correct but requires a minor modification to be fully usable or achieve the intended functionality.\\
        (0) Functionally Incorrect: The generated step differs in purpose, behavior, or misses key functionality.\\
        If uncertain, choose the most appropriate label.  
        Do not explain your reasoning. Output only the label number (e.g., 2).
        Now evaluate the following case:\\
        Input (Workflow Context):
        \{input\}\\
        Output (Generated Step):
        \{prediction\}\\
        Target (Ground Truth Step):
        \{target\}\\
        Your answer: <TO\_BE\_GENERATED>}

\end{comment}

\subsection{Result Analysis}

Experimental results show that all five LLMs successfully produced valid judgments for the workflow generation task, demonstrating that modern LLMs can understand and execute the evaluation prompt. However, understanding a task does not necessarily imply correct or meaningful judgments. Therefore, we further examine judgment quality by computing Krippendorff’s $\alpha$ between the LLM outputs and the human ground-truth, as shown in~\tabref{tab:judge_eval}.

\begin{table}[h]
\centering
\small
\setlength{\tabcolsep}{6pt}
\renewcommand{\arraystretch}{1.1}
\caption{Agreement, correlation, and API cost of LLM-as-a-Judge evaluators against human ground-truth}
\label{tab:judge_eval}
\begin{tabular}{l|c|c|c|c|c}
\toprule
\multirow{2}{*}{\textit{Evaluator}} 
& \multicolumn{2}{c|}{$\alpha$} 
& \multicolumn{2}{c|}{$\rho$} 
& \multirow{2}{*}{\textit{Cost}} \\
\cline{2-3}\cline{4-5}
& \textit{ZS} & \textit{OS} & \textit{ZS} & \textit{OS} & \rule{0pt}{2.4ex}\\
\midrule
Claude-Sonnet-4.5 & 0.648 & 0.656 & 0.648 & 0.658 & \$1.41 \\
Claude-Haiku-4.5  & 0.631 & 0.612 & 0.665 & 0.679 & \$0.47 \\
DeepSeek-V3.2-Exp & \textbf{0.701} & 0.655 & \textbf{0.703} & 0.663 & \textbf{\$0.07} \\
GPT-4.1           & 0.604 & \textbf{0.687} & 0.619 & \textbf{0.695} & \$0.74 \\
GPT-3.5-turbo     & 0.490 & 0.285 & 0.522 & 0.382 & \$0.37 \\
\midrule
BLEU              & -- & -- & 0.514 & 0.514 & -- \\
ROUGE             & -- & -- & 0.692 & 0.692 & -- \\
\bottomrule
\end{tabular}
\\[3pt]
\parbox{\linewidth}{\footnotesize * ZS: zero-shot; OS: one-shot. $\alpha$: Krippendorff's agreement with human ground-truth; $\rho$: Spearman correlation with human ground-truth. Cost is the average API expense of ZS and OS judging runs.}
\end{table}

\begin{figure*}[ht]

 \centering
     {\includegraphics[width=1\textwidth]{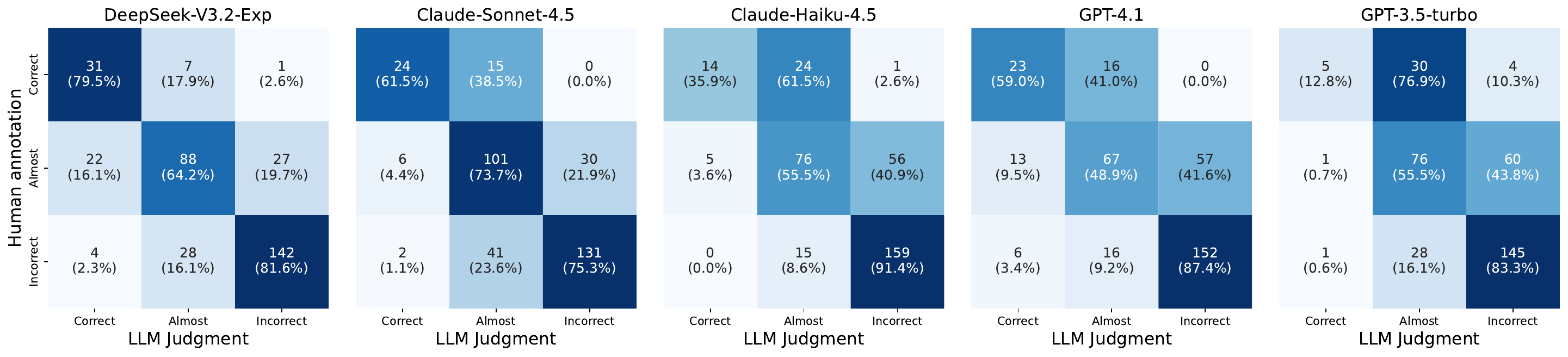}}
 \caption{Confusion matrices of LLM judges compared with human annotations under zero-shot prompting}
 \label{fig:confusion}

\end{figure*}

\paragraph{Human Alignment}  

%Recall that the two human annotators had $\alpha = 0.82$, serving as a practical benchmark for alignment with human judgment. Among the evaluated models, DeepSeek-V3.2-Exp shows the highest agreement ($\alpha=0.701$ zero-shot, $0.687$ one-shot), followed by Claude-Sonnet-4.5 ($0.648$–$0.656$) and Claude-Haiku-4.5 ($0.612$–$0.631$), all within the substantial agreement range. GPT-4.1 performs slightly worse ($0.604$–$0.635$), while GPT-3.5-turbo shows weak reliability ($0.285$–$0.490$). Unlike in \textit{Script} generation, where one-shot prompting benefits stronger models, it provides no consistent benefit for LLM-as-a-Judge and may even degrade weaker models, suggesting that a single example is insufficient to reliably guide LLM judges. Moreover, even the best LLM remains about 0.12 below human–human agreement ($0.70$ vs.\ $0.82$), indicating that although modern LLMs can approximate human evaluation reasonably well, they still fall short of human reliability and cannot yet fully replace human annotators in high-fidelity evaluation settings.

Recall that the two human annotators achieved $\alpha = 0.82$, serving as a practical benchmark for alignment with human judgment. DeepSeek-V3.2-Exp achieves the highest agreement in the zero-shot setting ($\alpha=0.701$), whereas GPT-4.1 performs best in the one-shot setting ($\alpha=0.687$). Claude-Sonnet-4.5 ($0.648$ zero-shot; $0.656$ one-shot) and Claude-Haiku-4.5 ($0.631$; $0.612$) also show substantial agreement. GPT-4.1 improves from $0.604$ to $0.687$ with one-shot prompting, while GPT-3.5-turbo exhibits fair-to-moderate agreement ($0.490$; $0.285$). Overall, one-shot prompting provides no consistent benefit across models: it improves agreement for Claude-Sonnet-4.5 and GPT-4.1 but reduces it for the other judges. Moreover, even the best LLM remains about 0.12 below human–human agreement ($0.70$ vs.\ $0.82$), indicating that although modern LLMs can approximate human evaluation reasonably well, they still fall short of human reliability and cannot yet fully replace human annotators in high-fidelity evaluation settings.

\paragraph{Comparison with Automatic Metrics}  
We further compare LLM-based judges with traditional automatic metrics (\ie BLEU and ROUGE). Since all 350 samples in this experiment correspond to incorrect script snippets, the accuracy is always zero and therefore not meaningful for comparison. To examine how well different evaluation approaches reflect human judgment, we compute the Spearman correlation $\rho$ between each metric and the human ground-truth. As shown in Table~\ref{tab:judge_eval}, BLEU achieves a moderate correlation with human labels ($\rho = 0.514$), while ROUGE reaches a stronger correlation ($\rho = 0.692$). The five LLM judges achieve correlations ranging from $0.382$ to $0.703$ across zero-shot and one-shot settings. In the zero-shot setting, DeepSeek-V3.2-Exp achieves the highest correlation ($\rho = 0.703$), while in the one-shot setting GPT-4.1 performs best ($\rho = 0.695$). These results suggest that the best-performing LLM-judge configurations match or slightly exceed widely used lexical-overlap metrics in their alignment with human judgment, while additionally assessing the functional correctness of generated scripts.

%These results suggest that LLM-based evaluation captures human judgment at least as well as widely used lexical-overlap metrics, while additionally providing an assessment of the functional correctness of the generated scripts.

\paragraph{Practical Evaluation Cost}
From a cost-efficiency perspective, the LLM judges also provide a practical advantage: in our experiment, judging all 350 samples cost only \$0.07--\$1.41 depending on the model, whereas expert annotation required substantial manual effort from multiple authors. This suggests that LLM-as-a-Judge can serve as an affordable first-pass evaluation mechanism, while human judgment remains necessary for calibration and adjudication.

\subsection{Discussion}

% To understand (1) why LLM judges do not reach the same level of agreement as human annotators and (2) why different LLM judges exhibit varying levels of alignment with human judgments, 
To investigate why LLM judges diverge from human consensus and exhibit inconsistent alignment across models, we analyze their prediction behavior using confusion matrices (\figref{fig:confusion}), using the zero-shot setting as an example. The results show that most disagreements occur in the \textit{Functionally Almost Correct} category, where borderline cases are frequently misclassified as either correct or incorrect, suggesting that LLM judges lack the sensitivity needed to reliably evaluate borderline functional equivalence. The matrices also reveal different evaluation tendencies across models. DeepSeek exhibits the most balanced predictions with most mass on the diagonal, which explains its highest correlation with human annotations. In contrast, Claude models and GPT-4.1 tend to be stricter by downgrading correct \textit{Scripts} to almost correct, while GPT-3.5 shows a strong conservative bias by rarely predicting correct samples. This suggests that LLM judges differ in both accuracy and strictness, meaning that model selection plays a critical role depending on whether conservative or balanced evaluation is desired.

\subsection{Implication}
Traditional lexical metrics such as BLEU and ROUGE, while useful for source code, often fail to capture the functional equivalence of generated scripts. 
Our findings indicate that LLM-based judges can bridge this ``semantic gap'' by evaluating the intent and logic of a generated \textit{Script} rather than its character-level overlap. 
% LLM-as-a-Judge can complement traditional metrics by providing semantic evaluation of generated \textit{Scripts}. 
While human expert annotation remains the ``gold standard'' for reliability, it is unscalable and expensive.
Although LLM judges do not yet match human reliability, their substantial alignment with human annotations suggests they can serve as a scalable and cost-effective evaluation method for workflow automation tasks.

%%%%%%%%%%%%%%%%%%%%%%%
\section{Threats to Validity}
\label{sec:threats}
%%%%%%%%%%%%%%%%%%%%%%%
%In this section, we discuss the threats to validity among our study and how we have mitigated them, 
% The following threats have been identified to validity:

\textbf{Threats to internal validity} concern whether the study accurately reflects cause–effect relationships. Since \approach consists of multiple models trained on large datasets, bias may arise from data preprocessing or training configurations (\eg random initialization, batch composition, and optimizer choice). To mitigate this, we fixed random seeds across experiments, used consistent data splits %, and repeated each experiment three times 
to support reproducibility. Another threat concerns the manual annotation in the LLM-as-a-Judge evaluation; to reduce subjective bias, two annotators independently labeled the samples, with disagreements resolved by a third annotator, and inter-rater agreement was measured before aggregating results.

%To mitigate this, we fixed random seeds across experiments, used consistent data splits for all models, and repeated each experiment three times to ensure stability. Another potential threat lies in the manual annotation used for the LLM-as-a-Judge evaluation; to reduce subjective bias, two annotators independently labeled samples, with a third annotator resolving complex conflicts, and inter-rater agreement was measured before aggregating results.

\textbf{Threats to external validity} concern the generalizability of our findings beyond the studied dataset and setup. \approach was evaluated on a large-scale dataset of GHA workflows from prior studies~\cite{actiondata,10.1145/3597503.3623351}. Although we refined this dataset (\eg separating \textit{Action} and \textit{Script} samples, filtering malformed workflows, and aligning context–step pairs), it still represents only open-source GitHub projects and may not reflect private or industrial workflows. Moreover, LLM-based evaluation depends on prompt design and model versions, which may yield different behaviors over time. To mitigate this, we used uniform prompting templates and documented all configurations to support replication.

\textbf{Threats to construct validity} concern whether our evaluation truly measures the intended concepts. Automatic metrics such as Top-K accuracy, BLEU-4, and ROUGE-L capture syntactic similarity but may overlook semantic equivalence. To address this, we adopted the LLM-as-a-Judge framework to evaluate correctness and contextual relevance, and analyzed inter-rater agreement among humans and between humans and LLMs to assess consistency. Another threat lies in the binary \textit{Action}–\textit{Script} distinction: some functionalities can be implemented either way, making the separation occasionally simplistic, which is a limitation to be explored in future work.

%\textbf{Threats to conclusion validity.}
%Conclusion validity refers to the soundness of our interpretations and statistical analyses. We used non-parametric statistical tests (e.g., Wilcoxon signed-rank) and effect-size measures to assess the significance of model differences, ensuring robustness against non-normal data distributions. Reported improvements were considered meaningful only when statistically significant across multiple runs. Nevertheless, differences in model size and training objectives might confound performance interpretation, so we cautiously framed conclusions in relative rather than absolute terms.
%%%%%%%%%%%%%%%%%%%%%%%
\section{Related work}
\label{sec:related}
%%%%%%%%%%%%%%%%%%%%%%%
%We present the existing studies concerning CI practices, and those explore the flaky behaviors in both CI process and tests. Researches on build outcome prediction are also presented.

\subsection{GitHub Actions and CI/CD Automation}
\label{subsec:action}

%Since its introduction, GHA has quickly become the dominant CI/CD service on GitHub, though research on it is still emerging. Golzadeh \etal~\cite{9825792} showed that GHA surpassed other CI tools within 18 months, while Kinsman \etal~\cite{9463074} found that its adoption changed collaboration dynamics yet was perceived positively by developers. %Valenzuela-Toledo and Bergel~\cite{valenzuela2022evolution} identified common workflow modification patterns and highlighted the need for better maintenance tools. 
%Decan \etal~\cite{actiondata} reported that \textit{Action} reuse is frequent but concentrated in a few popular \textit{Actions}. Bouzenia \etal~\cite{10.1145/3597503.3623303} studied resource usage in GHA by analyzing 1.3M runs, revealed high testing and build costs, and proposing optimizations for efficiency. Onsori Delicheh \etal~\cite{10.1145/3643991.3644899} found widespread dependency and input-validation vulnerabilities in reusable \textit{Actions}, and Zheng \etal~\cite{10.1145/3749371} categorized workflow failures, showing \textit{Scripts} are more error-prone than \textit{Actions}. A recent work~\cite{11025893} extended this line of research with a longitudinal security analysis of 18.9K workflows, showing improved permission practices but persistent risks such as secret leakage and unverified \textit{Actions}.

GHA has rapidly become a dominant CI/CD service on GitHub~\cite{9825792}, with studies showing broad developer adoption and generally positive perceptions~\cite{9463074}. Prior empirical work has characterized \textit{Action} reuse~\cite{actiondata}, resource consumption~\cite{10.1145/3597503.3623303}, dependency and input-validation vulnerabilities~\cite{10.1145/3643991.3644899}, workflow failures~\cite{10.1145/3749371}, and longitudinal security risks~\cite{11025893}. These studies establish the practical importance of GHA and motivate better authoring support for reliable workflow construction.

\subsection{Learning-based Program and Workflow Generation}

Learning-based techniques have also been applied to workflow-related prediction tasks. CIGAR~\cite{10356675} uses contrastive learning for \textit{Action} recommendation, while GH-WCOM~\cite{10.1145/3597503.3623351} applies T5 to workflow-step completion. Zhang \etal~\cite{10.1145/3664476.3664497} evaluated LLMs for workflow generation, defect detection, and repair, finding improvements in validity but remaining challenges in security and logical consistency. Gavel~\cite{10.1145/3674805.3690752} addresses a different task, categorizing reusable \textit{Actions} from README files to improve Marketplace visibility.

\subsection{LLMs for Software Engineering Tasks}

LLMs now support many SE tasks, including code generation, completion, summarization, bug localization, test generation, and automated repair~\cite{10.1145/3695988,chen2021evaluatinglargelanguagemodels,yin2017syntacticneuralmodelgeneralpurpose,yang2023assessingimprovingsyntacticadversarial,mastropaolo2022usingtransferlearningcoderelated,Vallecillos_Ruiz_2025,xia2023conversationalautomatedprogramrepair}. Related work also studies requirements engineering~\cite{HeyKeimKoziolek2020_1000124657,yang2022aspectbasedapireviewclassification}, prompt-engineering strategies~\cite{ahmed2024automaticsemanticaugmentationlanguage}, and practical developer use of tools such as ChatGPT and Copilot~\cite{10.1145/3643795.3648379}. More recently, Crupi \etal~\cite{crupi2025effectivenessllmasajudgecodegeneration} evaluated LLM-as-a-Judge for code generation and summarization, showing promise but also imperfect alignment with human judgment.

\subsection{Distinction and Contribution of FLOWSTEP}

Together, prior studies show the promise of learning-based workflow automation but leave two gaps. First, GH-WCOM~\cite{10.1145/3597503.3623351} treats workflow steps as monolithic outputs, while CIGAR~\cite{10356675} focuses only on \textit{Action} recommendation. Second, LLM-based workflow studies~\cite{10.1145/3664476.3664497} emphasize generation validity but provide limited treatment of contextual correctness and semantic evaluation. \approach addresses these gaps by explicitly modeling step type, predicting reusable \textit{Actions}, generating project-specific \textit{Scripts}, and evaluating generated scripts through human-calibrated LLM judges.

%%%%%%%%%%%%%%%%%%%%%%%
\section{Conclusions}
\label{sec:conclusions}
%%%%%%%%%%%%%%%%%%%%%%%

In this work, we introduced \approach, a framework designed to improve the recommendation of GitHub Actions workflow steps by explicitly modeling the heterogeneity between \textit{Actions} and \textit{Scripts}. By leveraging a ``divide-and-conquer'' architecture, we integrated a RoBERTa-based classifier for step-type prediction, contrastive learning for reusable \textit{Action} selection, and Large Language Models for project-specific \textit{Script} generation. Our evaluation demonstrates that this step-type-aware formulation substantially outperforms existing baselines.
As commonly used metrics do not fully capture the correctness of generated scripts, we further validated whether LLMs can serve as reliable evaluators. Our results show that there is substantial alignment between human annotations and LLMs' evaluations, indicating a promising path in this direction.

Although \approach improves workflow step generation accuracy, it relies solely on the workflow specification and does not exploit repository-specific information. As future work, we plan to investigate an agentic CI system that actively explores the target repository to leverage project-specific context for more accurate workflow generation and correction. We also plan to integrate vulnerability detection and security analysis to generate CI/CD workflows that are both functionally correct and secure by design.

%Since LLM-based evaluations and generation capabilities depend on specific model versions, future studies could investigate how the results may change with newer model releases over time. Moreover, while our approach improves accuracy, the framework could be extended to incorporate advanced features such as automated vulnerability detection and logical consistency checks to ensure the generation of more robust and secure CI/CD workflows.

% \section*{Data-Availability Statement} 

\textbf{Replication package.} To support replication studies and future extensions, the scripts and the dataset used in the study are available online at \url{https://github.com/jiangnanpro/FLOWSTEP}.

% \subsection{Replication Package}

\balance
% \newpage
% \IEEEtriggeratref{14}
\bibliographystyle{IEEEtran}
\bibliography{references}

\end{document}